\documentclass[reprint,amsmath,amssymb,aps,prl,superscriptaddress,reprint]{revtex4-2}

\makeatletter
\def\set@curr@file#1{%
  \begingroup
    \escapechar\m@ne
    \xdef\@curr@file{\expandafter\string\csname #1\endcsname}%
  \endgroup
}
\def\quote@name#1{"\quote@@name#1\@gobble""}
\def\quote@@name#1"{#1\quote@@name}
\def\unquote@name#1{\quote@@name#1\@gobble"}
\makeatother

\usepackage{graphicx}
\usepackage{dcolumn}
\usepackage{bm}
\usepackage{xcolor}

\graphicspath{{C:/Users/ale/Dropbox/lavoro/research/2021_BIC/manuscript/OM_BIC/figures/}}

\begin{document}

\title{Optomechanical Modulation Spectroscopy of Bound-States-In-The-Continuum in a dielectric metasurface}

\author{S. Zanotto}
\thanks{Equal contribution}
\affiliation{NEST Lab., CNR - Istituto di Nanoscienze and Scuola Normale Superiore, piazza San Silvestro 12, 56217 Pisa - Italy}

\author{G. Conte}
\thanks{Equal contribution}
\affiliation{NEST Lab., CNR - Istituto di Nanoscienze and Scuola Normale Superiore, piazza San Silvestro 12, 56217 Pisa - Italy}
\affiliation{Dipartimento di Fisica, Università di Pisa, Largo Pontecorvo 3, 56127 Pisa, Italy}

\author{L. C. Bellieres}
\affiliation{Nanophotonics Technology Center, Universitat Politècnica de Valencia, Spain }

\author{A. Griol}
\affiliation{Nanophotonics Technology Center, Universitat Politècnica de Valencia, Spain }

\author{Daniel Navarro-Urrios}
\affiliation{MIND-IN2UB, Departament d'Electrònica, Facultat de Física, Universitat de Barcelona, Martí i Franquès 1, 08028 Barcelona, Spain}

\author{A. Tredicucci}
\affiliation{NEST Lab., CNR - Istituto di Nanoscienze and Scuola Normale Superiore, piazza San Silvestro 12, 56217 Pisa - Italy}
\affiliation{Dipartimento di Fisica, Università di Pisa, Largo Pontecorvo 3, 56127 Pisa, Italy}

\author{A. Mart\'{i}nez}
\affiliation{Nanophotonics Technology Center, Universitat Politècnica de Valencia, Spain }

\author{A. Pitanti}
\affiliation{NEST Lab., CNR - Istituto di Nanoscienze and Scuola Normale Superiore, piazza San Silvestro 12, 56217 Pisa - Italy}

\begin{abstract}
Elusive features in photonic and electronic devices can be detected by means of advanced, time-domain spectroscopic techniques. In this letter we introduce a novel kind of modulation spectroscopy, based on the optomechanical interaction of photonic and mechanical modes. Applying the technique to a Si metasurface and its drum-like mechanical modes, we detect narrow-band quasi-Bound-State-in-the-Continuum (q-BIC) modes close to normal incidence, where their measurement can be hindered by a high symmetry protection and undesired background modes. Showing a visibility enhancement of more than one order of magnitude, the optomechanical modulation spectroscopy can be an innovative tool for precise spectroscopy of a wide set of photonic devices, including the goal of measuring purely symmetry protected BIC resonances.      
\end{abstract}

\maketitle
	
\textsl{Introduction} - Spectroscopic techniques are extremely powerful tools to investigate the properties of a wide set of materials and devices using almost non-invasive light probes \cite{Skoog2017}. Classical spectroscopy, where reflectivity and transmissivity of a test sample are investigated using tunable or broad-band continuous-wave (CW) sources, have been quickly sided by more complex techniques exploiting dynamical effects. Modulation-based spectroscopy can comprise either an \textit{external} modulation of the energy of the impinging monochromatic source (wavelength modulation spectroscopy) \cite{Werle1991,Bomse1992} or an \textit{inner} modulation of the sample properties. The latter approach manifests in many techniques, including excitation-modulation spectroscopy, photo-reflectance or pump-probe techniques in semiconductors \cite{Hexter1962,Cardona1970}; classic schemes have found a recent application in investigation of optical plasmons \cite{Takagi2017} and 2D materials \cite{Tolloczko2020}. Notably, a pure \textit{inner} modulation spectroscopy using single CW optical sources requires particular device characteristics: so far it essentially included the possibility of charge density tuning (charged modulation spectroscopy) in optoelectronic devices based on organic semiconductors \cite{Liu2019} and field-effect transistors \cite{Caironi2011}. Both \textit{outer} and \textit{inner} modulation techniques are based on the idea that the derivative of a weak signal can be used for peak detection \cite{VivoTruios2005} and for enhancing its visibility by background removal, within a scheme sometimes labeled derivative spectroscopy \cite{OHaver1982}.\\ 
\begin{figure}[t!]
\centering
\includegraphics[width=8cm]{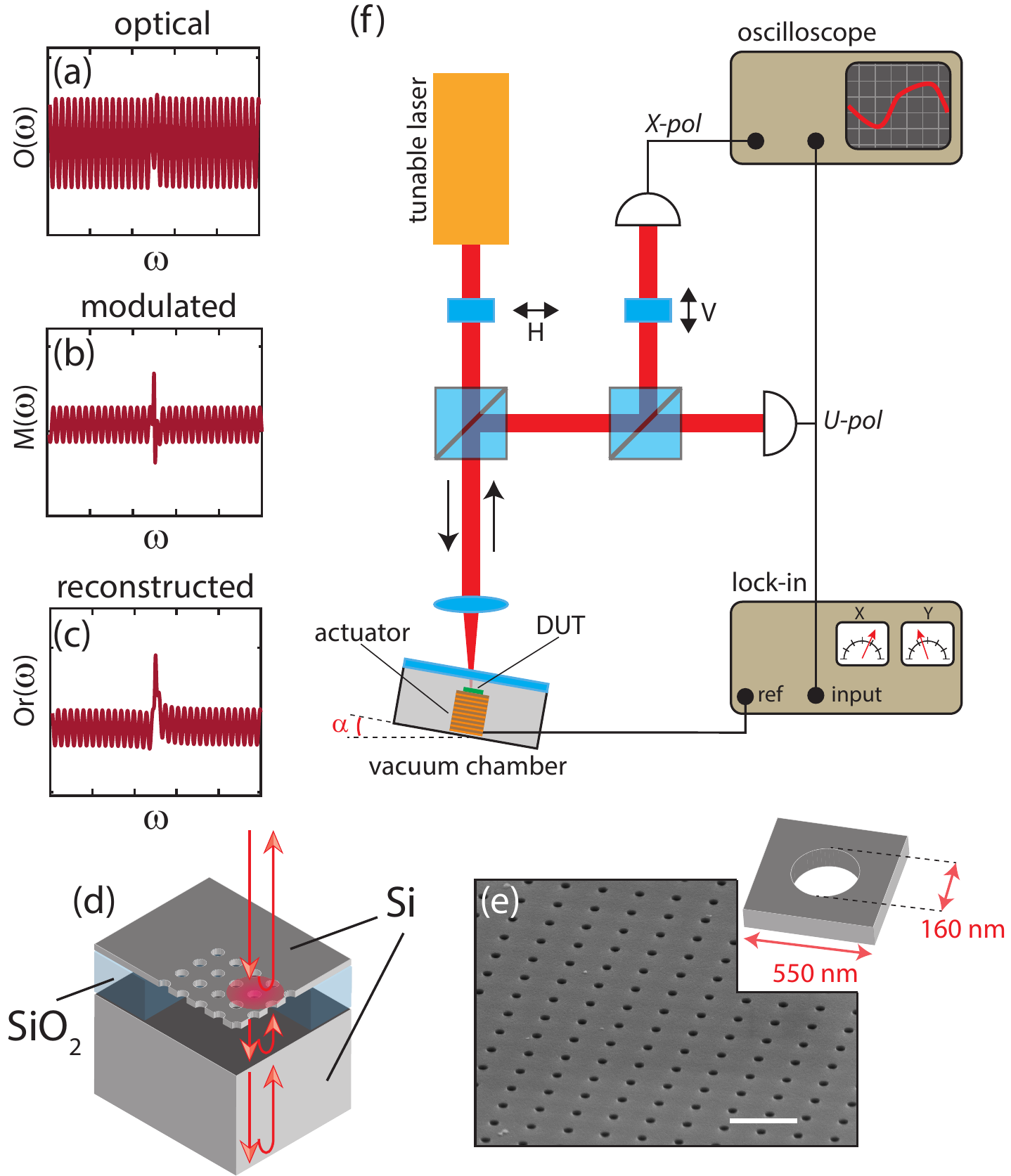}
\caption{OMS concept. Hidden features in the optical spectrum (a) are enhanced via the mechanically modulated signal (b), leading to the reconstructed optical signal (c). (d): Case study of a Si metasurface suspended on a Si handle, combining metasurface modes and multiple reflections within the substrate and air layer. (e): SEM micrograph of a fabricated device and sketch of the unit cell. White scale bar is 5 $\mu m$ (f): Sketch of the experimental setup for angle-resolved OMS.}
\label{fig:1}
\end{figure}
\begin{figure}[h!]
\centering
\includegraphics[width=8cm]{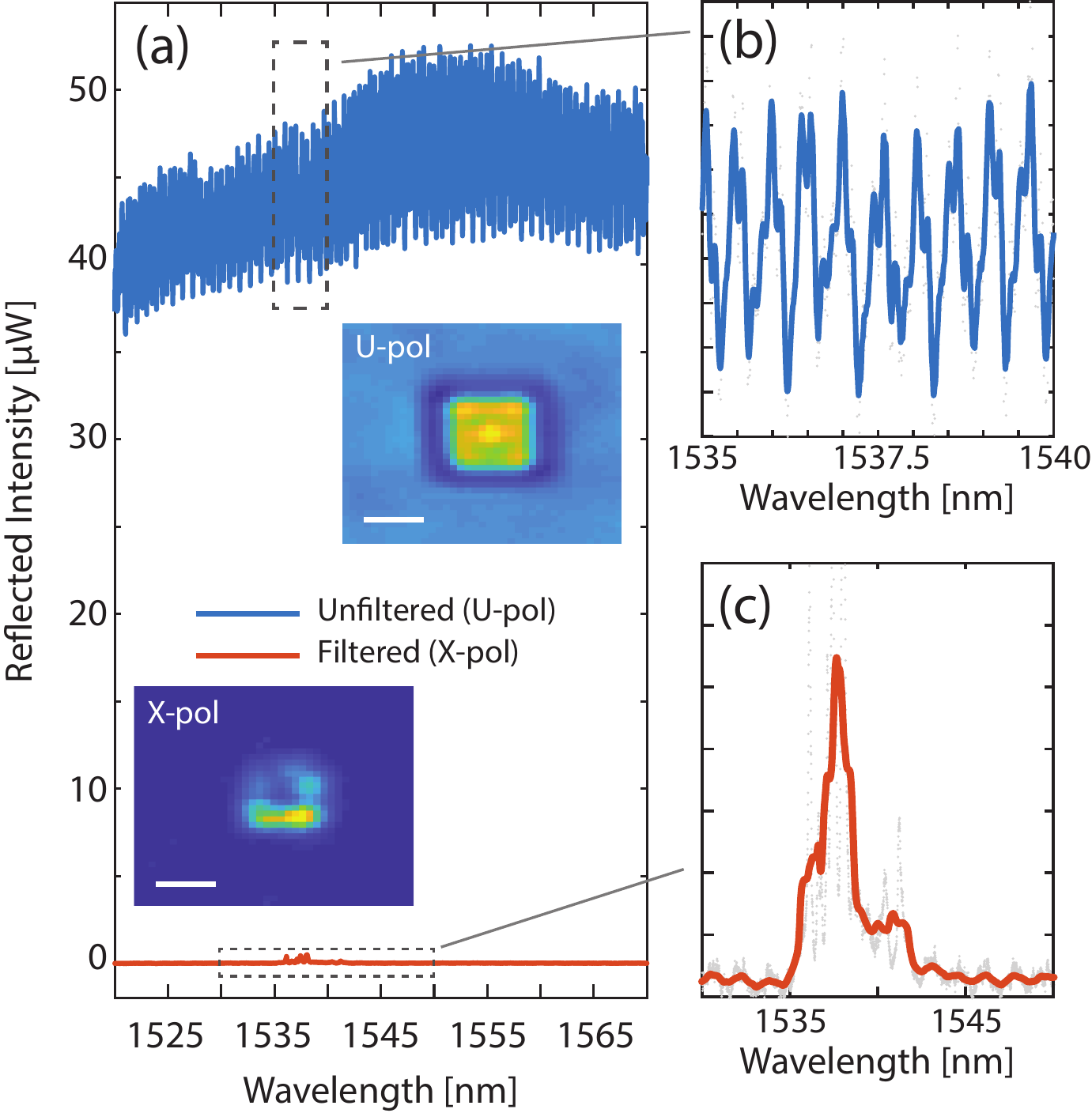}
\caption{Polarization dependent reflectivity response. (a) Unfiltered (U-pol) and cross-polarization filtered (X-pol) reflectivity signal from the device. A zoom-in of the two spectra are shown in (b) and (c), respectively. In the inset the maps of the average reflected intensity for U-pol and of the peak-to-peak reflected intensity for X-pol are shown.}
\label{fig:2}
\end{figure}
In this Letter, we introduce a novel kind of \textit{purely inner} spectroscopy technique, which we term Optomechanical Modulation Spectroscopy (OMS), based on optomechanical modulation of photonic device features. Optomechanics has recently risen as a research avenue where the coupling of photons with the ubiquitous mechanical modes has been investigated in several different kind of devices \cite{Aspelmeyer2014}. The foremost concept in this field is the possibility of modulating light through mechanical action, via a complex coefficient which governs dispersive and/or dissipative coupling. Remarkably, thermally-excited modes can modulate an incident optical wave without the need for generating external actuation. We anticipate that OMS can be used to selectively enhance photonic features in multi-resonance systems and can be implemented in all samples featuring mechanical action, which is pervasively present in photonic devices. For example, it can have an interest for photonic crystals, metasurfaces, whispering gallery mode resonators on top of substrates, as well as for fiber-coupled waveguide systems with undesired fiber-facet multiple reflection.
Here, we exploit the optomechanical modulation for enhancing spectroscopic features: the potential of OMS is shown by characterizing photonic Bound-States in the Continuum (BIC) in a Si metasurface. BICs are a general wave phenomenon where the states, despite having an energy within the radiative continuum, are interference- or symmetry-protected from leaking out to the external world. Being initially proposed in quantum mechanics \cite{Stilinger1975}, experimental and theoretical investigation of BICs has lately extended to several wave systems \cite{Hsu2016}, with a special focus on photonics, both in waveguide arrays \cite{Marinica2008, Plotnik2011} and planar systems \cite{Molina2012,Hsu2013,Hsu2013_2}. Pure BICs have infinite radiative Q-factors and therefore are invisible to external probes. The symmetry conditions necessary for a pure BIC can be fulfilled considering precise points in the system phase space; most commonly, BICs are found around $\Gamma$ points, yet in some cases accidental symmetry conditions could be realized in different points of the reciprocal space. Small deviations from the perfect symmetry protection couples the state with the continuum, giving tit a finite radiative Q-factor and granting the possibility of accessing it from outside (quasi-BIC). BICs and quasi-BICs have been subjected to a wide investigation, in the context of fundamental physics and topology \cite{Zhen2014, Doeleman2018} as well as for applications, offering ultra-high Q-factors in delocalized photonic devices such as metasurfaces. This allows exploiting far-field probing, granting several advantages to other ultra-high Q photonic systems which needs near-field coupling \cite{Akahane2003, Srinivasan2004,Armani2003}.
 
To understand the principle of OMS, we start by considering the optomechanical coupling $g_{\text{OM}}$ \cite{Aspelmeyer2014}, defined as:
\begin{equation}
g_{\text{OM}}=\frac{\partial\omega_0}{\partial x}
\label{eq:g}
\end{equation}  
where $\omega_0$ is the resonant frequency of a photonic mode and $x$ a generalized mechanical displacement. When a displacement $\Delta x$ is applied through mechanical motion, the static linear response optical signal from a resonant device, $O_0(\omega)$, gains an extra modulation term, $O(\omega) = O_0(\omega) + M(\omega) \Delta x$, where:
\begin{equation}
M(\omega)  =\frac{\partial O(\omega)}{\partial x} = - \frac{\partial O(\omega)}{\partial \omega}\cdot\frac{\partial\omega_0}{\partial x} = - O'(\omega)\cdot g_{\text{OM}},
\label{eq:O_M}
\end{equation}
more details can be found in the Supplementary Information (SI) - sec. I . On the right-hand side of eq. \ref{eq:O_M}, we get the first derivative of the optical signal; upon integration, we can then use the modulated signal to reconstruct the original optical signal, rescaled by the optomechanical coupling, $O_r(\omega)$. When several resonances are involved, we can use this technique to rescale each of them according to their own $g_{\text{OM}}$.
Figure \ref{fig:1} summarizes the main concept behind this idea. Let's suppose to have the optical feature of interest buried in a wide background (panel (a)). This feature could be a mode localized in a photonic structure, while the background could be coming from an underlying substrate. Moreover, let's suppose that the two contributions have different optomechanical couplings, the former possibly being larger than the latter, as routinely found in many photonic devices. Actuating a proper mechanical mode, the modulated signals can be evaluated (panel (b)), leading to a reconstructed signal (panel (c)), where the relative amplitude of the different features has been rescaled according to the ratio of their own optomechanical couplings, resulting in a strongly enhanced feature visibility of an elusive feature.\\

\textsl{Experimental results} - The BICs investigated with OMS have been implemented relying on a 220nm Si slab perforated with a square lattice of circular holes, more details on the design in SI - sec. II.A. The device layer, originally on top of sacrificial SiO$_2$, has been suspended, to maintain a local mirror symmetry along the vertical direction, leaving a two micrometers air layer between it and a thick Si wafer handle, see Fig. \ref{fig:1} (d). A SEM micrograph of a fabricated device along with a sketch of its unit cell are shown in Fig. \ref{fig:1} (e). More details on device fabrication can be found in the SI - sec. III. 

\begin{figure}[h!]
\centering
\includegraphics[width=8cm]{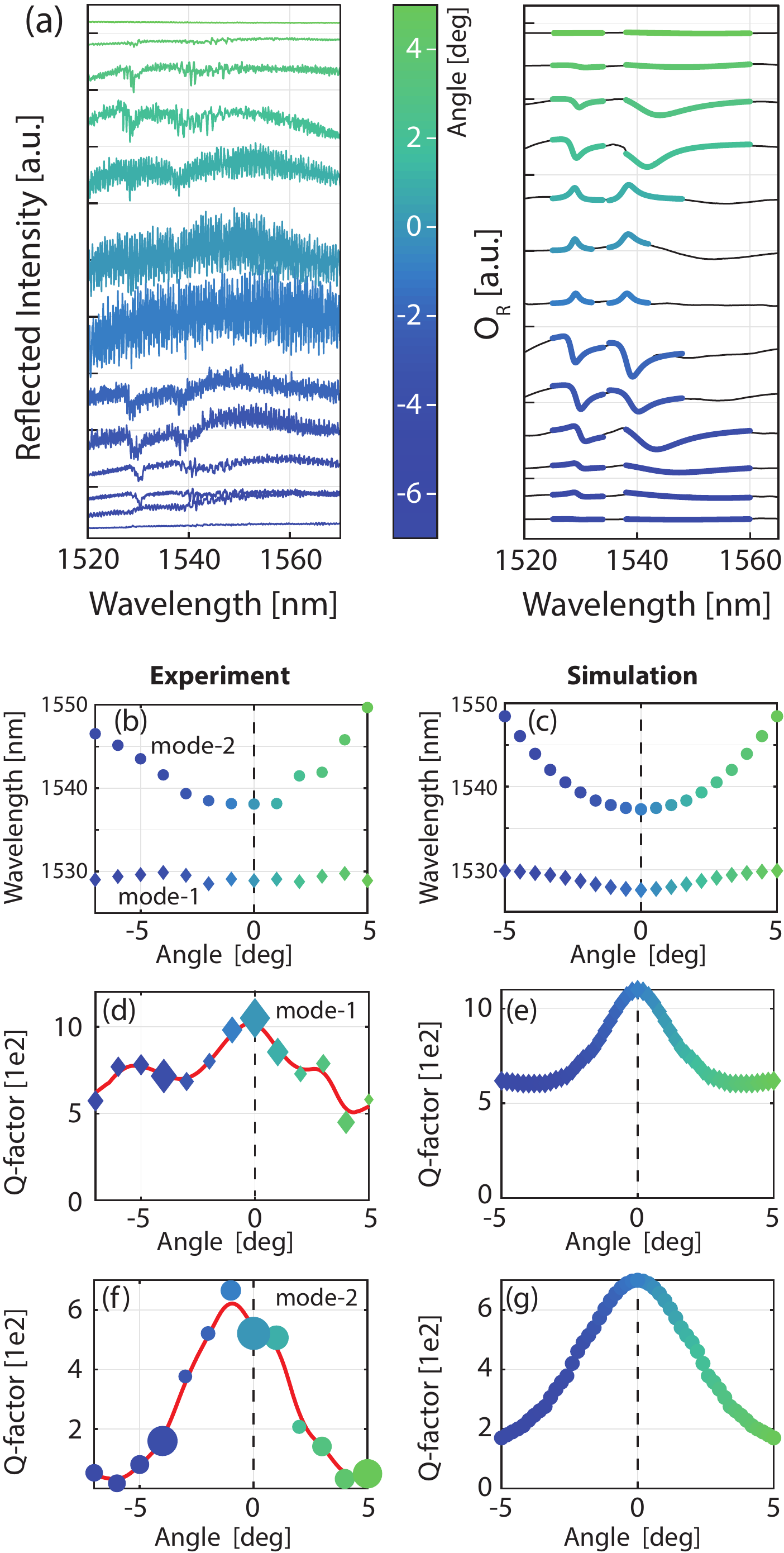}
\caption{q-BIC spectroscopy. (a): Waterfall plot of U-pol (left) and reconstructed (right) spectra at the metasurface center for different incident angles. Best fits with Fano lineshapes are indicated with bold, colored lines. (b)-(c): Experimental and simulated frequency dispersion. Experimental (d,f) and simulated Q-factors (e,g) for mode-1 and mode-2, respectively.}
\label{fig:4}
\end{figure}

Devices under test (DUTs) have been characterized using an angle-resolved reflectivity setup sketched in Fig. \ref{fig:1} (f). The setup is constituted by a tunable laser source (NewportVelocity TLB6700) whose 30 mW beam impinges on the DUT placed within a vacuum chamber on a motorized stage and with variable tilting $\alpha$. The filtered/unfiltered reflected signal intensity can be directly measured by means of an InGaAs photodiode (Newport 1623) using an oscilloscope (TekTronix MDO3104) or can be demodulated by a Lock-In amplifier (Zurich UHF). Additionally, a $f$=6 cm lens is used for beam focusing, while few optical components are needed to route the laser beam and manage its polarization state.  
Typical, unfiltered reflectivity data upon horizontally-polarized (H) laser illumination at normal incidence is shown in Fig. \ref{fig:2} (a) (U-pol). Here the signal is dominated by multiple reflections originating mostly from the substrate, appearing as narrowly-spaced intensity oscillations (see panel (b)). The average reflectivity is modified by the presence of the SiO$_2$ sacrificial layer; this can be used to spatially map the region where the metasurface has been defined. The figure inset shows a map of the average signal intensity, where the 75 $\mu m$ side metasurface can be clearly recognized. In this experimental configuration, the contribution from the metasurface modes is completely hidden and can be partially recovered only by filtering the reflected light using a linear polarizer with axis aligned to the vertical direction (V). The use of cross-polarized signals (X-pol, see Fig. \ref{fig:2} (a)) has been previously used in experiments involving photonic crystals with certain symmetries \cite{Galli2009,Leijssen2015}; in our case, the X-pol component originates, at normal incidence, from imperfections in the metasurface pattern (hole circularity, etc.) as well as local membrane bending. Since the polarization rotation power is weaker/absent in the device substrate, X-pol measurement could be used for enhancing the metasurface modes, as shown in the zoom-in of Fig. \ref{fig:2} (c). As a drawback, the X-pol signal is spatially inhomogeneous, since it depends on the local membrane structure. For example, X-pol peak-to-peak map (inset of Fig. \ref{fig:2} (a)) shows large values upon a metasurface corner, while being weaker on the opposite one. This fact, together with the different sensitivities of different modes to local inhomogeneities and their weak intensity makes the use of X-pol spectra not always reliable for optical spectroscopy. SI - sec. IV.B reports an extensive analysis of the X-pol spectra.\\
\begin{figure}[t!]
\centering
\includegraphics[width=8cm]{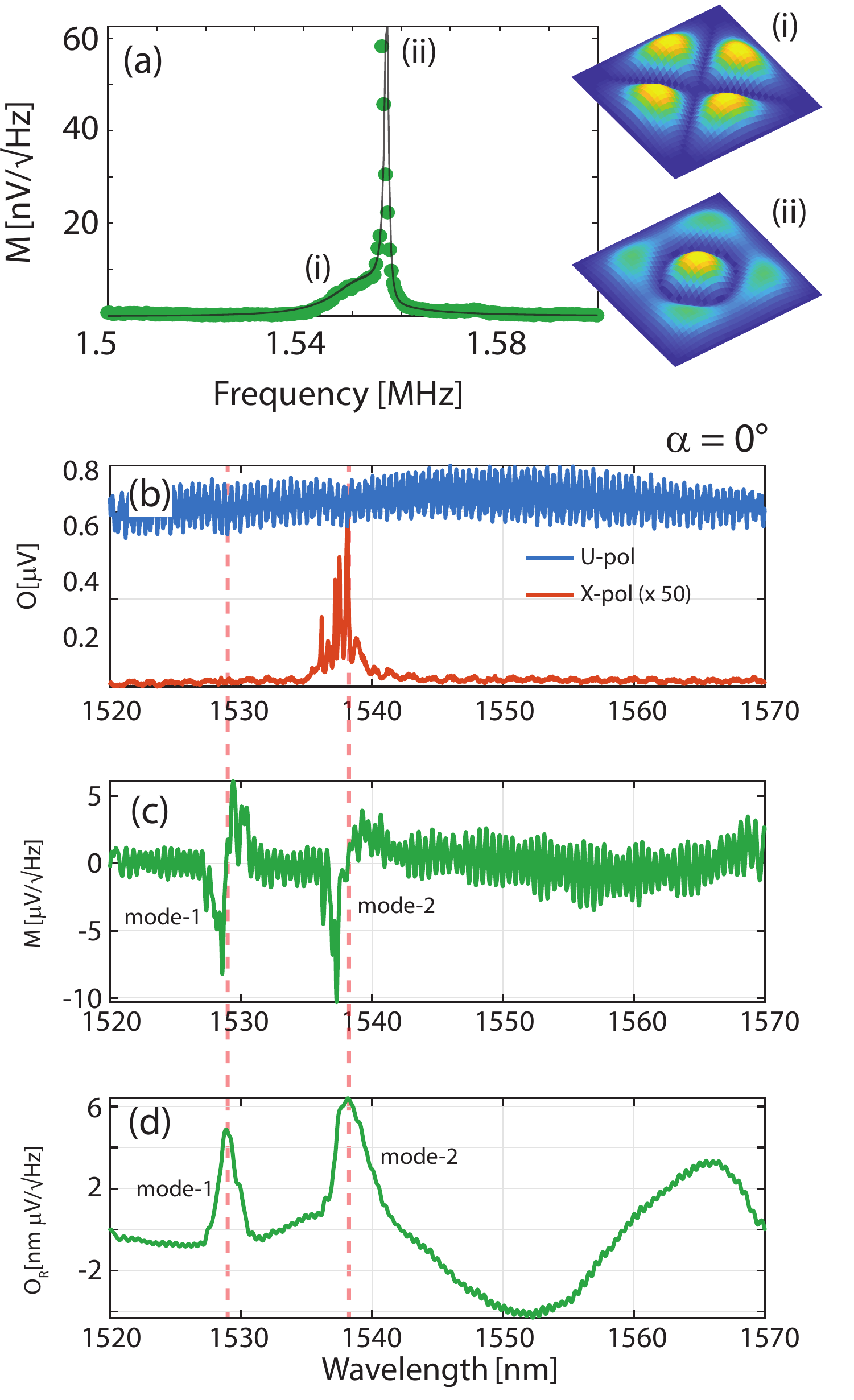}
\caption{Shaking the q-BIC. (a): Mechanical mode used for OMS. (b): Optical U-pol and X-pol spectra around normal incidence. (c): Demodulated modulation intensity. (d): reconstructed signal. The dashed pink lines indicate the spectral position of the q-BICs under investigation.}
\label{fig:3}
\end{figure}
OMS can be introduced by exciting membrane mechanical modes; this is done by placing the device on a piezoelectric piezoceramic multilayer actuator and operating in mild vacuum ($\sim 10^{-3}$ mbar). Decreasing the thin film damping, few membrane mechanical modes can be detected by demodulating the polarization-unfiltered light \footnote{The vanishing intensity of the X-pol signal prevents its direct demodulation at the LIA, forcing the use of the stronger U-pol signal.} reflected from the device with the AC bias fed to the piezo using the Lock-In amplifier (see Fig. \ref{fig:1} (f)) \cite{Baldacci2016}. The mechanical mode that we selected is shown in Fig. \ref{fig:3} (a); this has been chosen since it is well separated from the strongest resonance frequencies intrinsic to the piezo actuator (see SI - sec. IV.A). Note that OMS does not directly depend on the nature of the mechanical mode employed as long as it gives a stronger coupling with the mode of interest with respect to the undesired ones, which is a common configuration when localized and delocalized resonances are considered. The mechanical feature is composed by two peaks, holding Q-factors of about 100 and 1500, respectively, limited by molecular damping \cite{Baldacci2016,Zanotto2019} \footnote{Q-factors are limited by molecular film damping, although their value do not directly impact the performance of OMS as long as the resonator is not overdamped (like at atmospheric pressure) and its spectral features can be clearly recognized.}. The finite-element method (FEM) simulated modal shape attributed to both modes has been reported in the same figure. These simulations have shown a large degree of reliability when used in different photonic devices implemented in the same material platform \cite{GomisBresco2014}.           	
Feeding the piezo with an AC tone at a frequency of 1.557 MHz, corresponding to the maximum of mode (ii) peak, we swept the laser wavelength and recorded the reflected and the demodulated intensity (O and M, respectively). Figure \ref{fig:3} (b) reports the U-pol and X-pol optical spectra, the latter having been rescaled for visibility. A single narrow feature can be identified only in the X-pol spectra, whereas the modulation amplitude, $M$, clearly shows two narrow features, one around 1528 nm (mode-1) and the other around 1537 nm (mode-2), respectively, see Fig. \ref{fig:3} (c). The reconstructed optical signal, shown in panel (d), has now two clear resonant features, with a large enhancement in the visibility with respect to the original U-pol reflectivity data\footnote{To improve the visibilities at large angles, a complementary dataset to the one of Fig. \ref{fig:4}, with individually normalized spectra, has been reported in the SI - sec. IV.B.}. Interestingly, using OMS, we detected two resonant peaks within the laser wavelength range, one of which was completely absent even in the X-pol spectrum. The different X-pol intensity of the two modes is linked to their different sensitivity to imperfections: the electromagnetic field of mode-1 is less confined around the circular metasurface hole than mode-2, the latter being more affected by symmetry breaking involving the hole circularity or surface roughness, see SI - sec. II.A. Inspecting eq. (\ref{eq:O_M}), we expect the mode enhancement to be proportional to each mode $g_{\text{OM}}$. Using rigorous coupled wave method based simulations \cite{PPML}, we found, at almost normal incidence and in the membrane center, a $g_{\text{OM}}^s=2.99 \text{ kHz}$ optomechanical coupling for the substrate modes and $g_{\text{OM}}^{m1}=51.05 \text{ kHz}$ and $g_{\text{OM}}^{m2}=74.4 \text{ kHz}$ for mode-1 and 2, respectively. These are compatible with the contrast enhancement in Fig. \ref{fig:3} (c), with a relative increase of about a factor 17 and 25 for mode-1 and mode-2, respectively. More details on the simulations can be found in the SI - sec. II.B.   
To show the power of OMS in the investigation of BICs, we performed standard and optomechanical modulation spectroscopy of the device at different angle of incidence $\alpha$, illuminating the device in the center of the metasurface. Figure \ref{fig:4} reports the main results where different spectra have been vertically shifted for clarity \footnote{For improved clarity of the waterfall plot, each curve has bee offset by a factor $c\cdot \tan^{-1}(10\alpha)$ , where $c$ is a constant and $\alpha$ the angle of incidence} and the angle of incidence color-coded from blue to green. In the U-pol spectra (leftmost panel) the two q-BIC features are visible at large angles, while they are completely hidden by the substrate contribution around normal incidence. This contrast difference can be expected from a simple model based on coupled-mode theory for resonances with varying radiative loss and a constant, small non-radiative loss contribution \cite{Zanotto2018}; detailed considerations can be found in the SI - sec. II.C. Even if using tilted U-pol spectroscopy for identifying q-BICs at finite angles, its performance degrades when getting closer to normal incidence, which is the region with the most detrimental substrate effect yet, conversely, the most interesting one since the q-BICs tends to the pure BIC modes, see SI - sec. IV.C. A very different situation can be found for the reconstructed signals in the rightmost panel: both q-BICs can be identified with a large signal-to-background ratio at every angle, with a generally improved visibility over the U-pol measurement. All the reconstructed signals have been fitted employing Fano functions, which are naturally used in situation where multiple resonances with different linewidths are coupled together \cite{Fan2002,Fan2003}. Figure \ref{fig:4} (b) shows the extracted resonant wavelength for mode-1 and mode-2 (diamonds and circles, respectively). As can be seen, mode-1 has an almost flat dispersion in the investigated region whereas mode-2 has a parabolic one. The resonant wavelengths are in very good agreement with FEM simulations for the infinite photonic crystal, which are reported in Fig. \ref{fig:4} (c). In q-BIC, it is expected that the optical Q-factors degrade moving far from the high symmetry configuration at $\alpha=0$; this can be see in the Q-factors obtained for mode-1 and mode-2 shown in panel (d) and (f), respectively. Here the scatter symbol size is directly proportional to the feature visibility, in such a way to elucidate the effect of OMS especially around normal incidence. Large Qs of about 1000 and 650 at $\alpha=0$ reduce to about 500 and 100 at large angles, respectively. Interesting, mode-2 shows a more marked decrease, whereas mode-1 tends to a constant value in the investigated range. In the ideal metasurface simulations, the imaginary part of the simulated eigenfrequencies is exclusively due to the radiative loss channel; this results in a radiative Q-factor, $Q_r$, diverging at normal incidence. For a more realistic comparison, we artificially included the effects of non-ideal patterning by rescaling the simulated $Q_r$ by an empirical factor $\alpha_{nr}=20$. Moreover, we model the effect of different loss channels (roughness-induced scattering, surface state absorption etc.) not included in the simulation by introducing an extra inner loss term, $Q_m$, which has been taken as $Q_m=1100$ and $Q_m=700$ for mode-1 and mode-2 respectively. The resulting total Q-factor, which can be written as: $Q_{tot}^{-1}=(Q_r/\alpha_{nr})^{-1}+Q_m^{-1}$, is reported in Fig. \ref{fig:4} (e) and (g): the semi-empirical estimation shows a good agreement with the outcome of experimental fits. For the sake of comparison, we tried to analyzed the spectra measured in X-pol; while obtaining acceptable results for mode-2, the weaker mode-1 cannot be reliably fitted, especially in the region around $\alpha=0$, where q-BIC shows their most interesting regime. The full data analysis can be found in the SI - sec. IV.C. Note that, while U-pol and X-pol tilted spectroscopy are valid tools for identifying q-BICs at finite angles, both signals do not produce clear results when normal incidence is evaluated, which in our case is the most interesting point since it represents the condition closer to the pure BIC modes, see SI - sec. IV.C.   

\textsl{Conclusions} - In this Letter we have proposed and demonstrated the application of a new class of derivative spectroscopy - Optomechanical Modulation Spectroscopy - to the q-BIC modes of a Si metasurface. The powerful technique exploits the mechanically induced modulation to analyze weak or hidden spectral features giving them a contrast proportional to the ratio of their optomechanical coupling divided by the masking background one. The range of application includes a large ensemble of photonic devices sitting on top of a substrate/handle layer: localized optical resonances can be enhanced by actuating the proper mechanical mode, which could be a breathing mode in non-suspended devices. Combining q-BIC resonances with mechanical resonators can be an interesting avenue for dynamical control of narrow photonic features: on one side, the high Q-factor could be used to exploit radiation pressure for optomechanical effects in metasurfaces and, more in general, in devices with delocalized photonic resonances where polarization effects can be better exploited \cite{Zanotto2019}. On the other, the mechanical perturbation could accurately control the BIC to q-BIC transition in ultra high-quality devices, allowing working at normal incidence for applications in the guise of what is done with dark and bright excitons in light-matter systems \cite{Ohta2018}.        

\textsl{Conclusions} - In this Letter we have proposed and shown the application of a new class of derivative spectroscopy - Optomechanical Modulation Spectroscopy - which exploits a mechanical modulation to analyze weak or hidden spectral features. It enables the use of derivative spectroscopy through inner modulation and with CW single sources, enhancing selected features in multi-resonance systems. Given the pervasive nature of flexural, torsional or breathing mechanical modes, OMS founds applications in a wide class of photonic devices, especially when undesirable background signals originating for the substrate/coupling scheme masks the resonances under investigation.\\ 
OMS has been validated investigating an optomechanical BIC in a dielectric metasurface, which is a hybrid device here experimentally characterized for the fist time. In perspective, such platform can be an interesting avenue for dynamical control of narrow photonic features: on one side, the high Q-factor could be used to exploit radiation pressure for optomechanical effects in metasurfaces and, more in general, in devices with delocalized photonic resonances where polarization effects can be better exploited \cite{Zanotto2019}. On the other, the mechanical perturbation could be used to accurately control the BIC to q-BIC transition in future ultra high-quality photonic devices, allowing working at normal incidence for applications in the guise of what is done with dark and bright excitons in light-matter systems \cite{Ohta2018}.    

\section{Acknowledgments}
A. M. acknowledges Funding from Ministerio de Ciencia, Innovación y Universidades (PGC2018-094490-B-C21, PRX18/00126) and Generalitat Valenciana (PROMETEO/2019/123, IDIFEDER/2021/061)

\bibliography{OM_BIC}

\begin{thebibliography}{38}%
\makeatletter
\providecommand \@ifxundefined [1]{%
 \@ifx{#1\undefined}
}%
\providecommand \@ifnum [1]{%
 \ifnum #1\expandafter \@firstoftwo
 \else \expandafter \@secondoftwo
 \fi
}%
\providecommand \@ifx [1]{%
 \ifx #1\expandafter \@firstoftwo
 \else \expandafter \@secondoftwo
 \fi
}%
\providecommand \natexlab [1]{#1}%
\providecommand \enquote  [1]{``#1''}%
\providecommand \bibnamefont  [1]{#1}%
\providecommand \bibfnamefont [1]{#1}%
\providecommand \citenamefont [1]{#1}%
\providecommand \href@noop [0]{\@secondoftwo}%
\providecommand \href [0]{\begingroup \@sanitize@url \@href}%
\providecommand \@href[1]{\@@startlink{#1}\@@href}%
\providecommand \@@href[1]{\endgroup#1\@@endlink}%
\providecommand \@sanitize@url [0]{\catcode `\\12\catcode `\$12\catcode
  `\&12\catcode `\#12\catcode `\^12\catcode `\_12\catcode `\%12\relax}%
\providecommand \@@startlink[1]{}%
\providecommand \@@endlink[0]{}%
\providecommand \url  [0]{\begingroup\@sanitize@url \@url }%
\providecommand \@url [1]{\endgroup\@href {#1}{\urlprefix }}%
\providecommand \urlprefix  [0]{URL }%
\providecommand \Eprint [0]{\href }%
\providecommand \doibase [0]{https://doi.org/}%
\providecommand \selectlanguage [0]{\@gobble}%
\providecommand \bibinfo  [0]{\@secondoftwo}%
\providecommand \bibfield  [0]{\@secondoftwo}%
\providecommand \translation [1]{[#1]}%
\providecommand \BibitemOpen [0]{}%
\providecommand \bibitemStop [0]{}%
\providecommand \bibitemNoStop [0]{.\EOS\space}%
\providecommand \EOS [0]{\spacefactor3000\relax}%
\providecommand \BibitemShut  [1]{\csname bibitem#1\endcsname}%
\let\auto@bib@innerbib\@empty
\bibitem [{\citenamefont {Skoog}\ \emph {et~al.}(2017)\citenamefont {Skoog},
  \citenamefont {Holler},\ and\ \citenamefont {Crouch}}]{Skoog2017}%
  \BibitemOpen
  \bibfield  {author} {\bibinfo {author} {\bibfnamefont {D.~A.}\ \bibnamefont
  {Skoog}}, \bibinfo {author} {\bibfnamefont {F.~J.}\ \bibnamefont {Holler}},\
  and\ \bibinfo {author} {\bibfnamefont {S.~R.}\ \bibnamefont {Crouch}},\
  }\href@noop {} {\emph {\bibinfo {title} {Principles of Instrumental
  Analysis}}}\ (\bibinfo  {publisher} {Cencage, Boston MA},\ \bibinfo {year}
  {2017})\BibitemShut {NoStop}%
\bibitem [{\citenamefont {Werle}\ and\ \citenamefont
  {Slemr}(1991)}]{Werle1991}%
  \BibitemOpen
  \bibfield  {author} {\bibinfo {author} {\bibfnamefont {P.}~\bibnamefont
  {Werle}}\ and\ \bibinfo {author} {\bibfnamefont {F.}~\bibnamefont {Slemr}},\
  }\bibfield  {title} {\bibinfo {title} {{Signal-to-noise ration analysis in
  laser absorption sspectrometers using optical multipass cells}},\ }\href@noop
  {} {\bibfield  {journal} {\bibinfo  {journal} {Appl. Opt.}\ }\textbf
  {\bibinfo {volume} {30}},\ \bibinfo {pages} {430} (\bibinfo {year}
  {1991})}\BibitemShut {NoStop}%
\bibitem [{\citenamefont {Bomse}\ \emph {et~al.}(1991)\citenamefont {Bomse},
  \citenamefont {Stanton},\ and\ \citenamefont {Silver}}]{Bomse1992}%
  \BibitemOpen
  \bibfield  {author} {\bibinfo {author} {\bibfnamefont {D.~S.}\ \bibnamefont
  {Bomse}}, \bibinfo {author} {\bibfnamefont {A.~C.}\ \bibnamefont {Stanton}},\
  and\ \bibinfo {author} {\bibfnamefont {J.~A.}\ \bibnamefont {Silver}},\
  }\bibfield  {title} {\bibinfo {title} {{Frequency modulation and wavelength
  modulation spectroscopies: comparison of experimental methods using a
  lead-salt diode laser}},\ }\href@noop {} {\bibfield  {journal} {\bibinfo
  {journal} {Appl. Opt.}\ }\textbf {\bibinfo {volume} {31}},\ \bibinfo {pages}
  {718} (\bibinfo {year} {1991})}\BibitemShut {NoStop}%
\bibitem [{\citenamefont {Hexter}(1963)}]{Hexter1962}%
  \BibitemOpen
  \bibfield  {author} {\bibinfo {author} {\bibfnamefont {R.~M.}\ \bibnamefont
  {Hexter}},\ }\bibfield  {title} {\bibinfo {title} {{Excitation-Modulation
  Spectroscopy: A Technique for Obtaining Vibrational Spectra of Excited
  Electronic States}},\ }\href@noop {} {\bibfield  {journal} {\bibinfo
  {journal} {J. Opt. Soc. Am.}\ }\textbf {\bibinfo {volume} {53}},\ \bibinfo
  {pages} {703} (\bibinfo {year} {1963})}\BibitemShut {NoStop}%
\bibitem [{\citenamefont {Cardona}(1970)}]{Cardona1970}%
  \BibitemOpen
  \bibfield  {author} {\bibinfo {author} {\bibfnamefont {M.}~\bibnamefont
  {Cardona}},\ }\bibfield  {title} {\bibinfo {title} {Modulation spectroscopy
  of semiconductors},\ }in\ \href@noop {} {\emph {\bibinfo {booktitle}
  {Festkörperprobleme 10. Advances in Solid State Physics, vol 10}}},\
  \bibinfo {editor} {edited by\ \bibinfo {editor} {\bibfnamefont
  {O.}~\bibnamefont {Madelung}}}\ (\bibinfo  {publisher} {Springer, Berlin},\
  \bibinfo {year} {1970})\BibitemShut {NoStop}%
\bibitem [{\citenamefont {Takagi}\ \emph {et~al.}(2017)\citenamefont {Takagi},
  \citenamefont {Nair}, \citenamefont {Saito}, \citenamefont {Seto},
  \citenamefont {Watanabe}, \citenamefont {Kobayashi},\ and\ \citenamefont
  {Tokunaga}}]{Takagi2017}%
  \BibitemOpen
  \bibfield  {author} {\bibinfo {author} {\bibfnamefont {K.}~\bibnamefont
  {Takagi}}, \bibinfo {author} {\bibfnamefont {S.~V.}\ \bibnamefont {Nair}},
  \bibinfo {author} {\bibfnamefont {J.}~\bibnamefont {Saito}}, \bibinfo
  {author} {\bibfnamefont {K.}~\bibnamefont {Seto}}, \bibinfo {author}
  {\bibfnamefont {R.}~\bibnamefont {Watanabe}}, \bibinfo {author}
  {\bibfnamefont {T.}~\bibnamefont {Kobayashi}},\ and\ \bibinfo {author}
  {\bibfnamefont {E.}~\bibnamefont {Tokunaga}},\ }\bibfield  {title} {\bibinfo
  {title} {{Plasmon Modulation Spectroscopy of Noble Metals to Reveal the
  Distribution of the Fermi Surface Electrons in the Conduction Band}},\
  }\href@noop {} {\bibfield  {journal} {\bibinfo  {journal} {Appl. Sci.}\
  }\textbf {\bibinfo {volume} {7}},\ \bibinfo {pages} {1315} (\bibinfo {year}
  {2017})}\BibitemShut {NoStop}%
\bibitem [{\citenamefont {oczko}\ \emph {et~al.}(2020)\citenamefont {oczko},
  \citenamefont {Oliva}, \citenamefont {Woźniak}, \citenamefont {Kopaczek},
  \citenamefont {Scharoch},\ and\ \citenamefont {Kudrawiec}}]{Tolloczko2020}%
  \BibitemOpen
  \bibfield  {author} {\bibinfo {author} {\bibfnamefont {A.~T.}\ \bibnamefont
  {oczko}}, \bibinfo {author} {\bibfnamefont {R.}~\bibnamefont {Oliva}},
  \bibinfo {author} {\bibfnamefont {T.}~\bibnamefont {Woźniak}}, \bibinfo
  {author} {\bibfnamefont {J.}~\bibnamefont {Kopaczek}}, \bibinfo {author}
  {\bibfnamefont {P.}~\bibnamefont {Scharoch}},\ and\ \bibinfo {author}
  {\bibfnamefont {R.}~\bibnamefont {Kudrawiec}},\ }\bibfield  {title} {\bibinfo
  {title} {{Anisotropic optical properties of GeS investigated by optical
  absorption and photoreflectance}},\ }\href@noop {} {\bibfield  {journal}
  {\bibinfo  {journal} {Mater. Adv.}\ }\textbf {\bibinfo {volume} {1}},\
  \bibinfo {pages} {1886} (\bibinfo {year} {2020})}\BibitemShut {NoStop}%
\bibitem [{\citenamefont {Liu}\ \emph {et~al.}(2019)\citenamefont {Liu},
  \citenamefont {Foo}, \citenamefont {Zapien}, \citenamefont {Li},\ and\
  \citenamefont {Tsang}}]{Liu2019}%
  \BibitemOpen
  \bibfield  {author} {\bibinfo {author} {\bibfnamefont {T.}~\bibnamefont
  {Liu}}, \bibinfo {author} {\bibfnamefont {Y.}~\bibnamefont {Foo}}, \bibinfo
  {author} {\bibfnamefont {J.~A.}\ \bibnamefont {Zapien}}, \bibinfo {author}
  {\bibfnamefont {M.}~\bibnamefont {Li}},\ and\ \bibinfo {author}
  {\bibfnamefont {S.-W.}\ \bibnamefont {Tsang}},\ }\bibfield  {title} {\bibinfo
  {title} {{A generalized Stark effect electromodulation model for extracting
  excitonic properties in organic semiconductors}},\ }\href@noop {} {\bibfield
  {journal} {\bibinfo  {journal} {Nat. Comm.}\ }\textbf {\bibinfo {volume}
  {10}},\ \bibinfo {pages} {5089} (\bibinfo {year} {2019})}\BibitemShut
  {NoStop}%
\bibitem [{\citenamefont {Caironi}\ \emph {et~al.}(2011)\citenamefont
  {Caironi}, \citenamefont {Bird}, \citenamefont {Fazzi}, \citenamefont {Chen},
  \citenamefont {Pietro}, \citenamefont {Newman}, \citenamefont {Facchetti},\
  and\ \citenamefont {Sirringhaus}}]{Caironi2011}%
  \BibitemOpen
  \bibfield  {author} {\bibinfo {author} {\bibfnamefont {M.}~\bibnamefont
  {Caironi}}, \bibinfo {author} {\bibfnamefont {M.}~\bibnamefont {Bird}},
  \bibinfo {author} {\bibfnamefont {D.}~\bibnamefont {Fazzi}}, \bibinfo
  {author} {\bibfnamefont {Z.}~\bibnamefont {Chen}}, \bibinfo {author}
  {\bibfnamefont {R.~D.}\ \bibnamefont {Pietro}}, \bibinfo {author}
  {\bibfnamefont {C.}~\bibnamefont {Newman}}, \bibinfo {author} {\bibfnamefont
  {A.}~\bibnamefont {Facchetti}},\ and\ \bibinfo {author} {\bibfnamefont
  {H.}~\bibnamefont {Sirringhaus}},\ }\bibfield  {title} {\bibinfo {title}
  {{Very Low Degree of Energetic Disorder as the Origin of High Mobility in an
  n-channel Polymer Semiconductor}},\ }\href@noop {} {\bibfield  {journal}
  {\bibinfo  {journal} {Adv. Funct. Mat.}\ }\textbf {\bibinfo {volume} {21}},\
  \bibinfo {pages} {3371} (\bibinfo {year} {2011})}\BibitemShut {NoStop}%
\bibitem [{\citenamefont {Vivó-Truyols}\ \emph {et~al.}(2005)\citenamefont
  {Vivó-Truyols}, \citenamefont {Torres-Lapasió}, \citenamefont {van
  Nederkassel}, \citenamefont {Heyden},\ and\ \citenamefont
  {Massart}}]{VivoTruios2005}%
  \BibitemOpen
  \bibfield  {author} {\bibinfo {author} {\bibfnamefont {G.}~\bibnamefont
  {Vivó-Truyols}}, \bibinfo {author} {\bibfnamefont {J.}~\bibnamefont
  {Torres-Lapasió}}, \bibinfo {author} {\bibfnamefont {A.}~\bibnamefont {van
  Nederkassel}}, \bibinfo {author} {\bibfnamefont {Y.~V.}\ \bibnamefont
  {Heyden}},\ and\ \bibinfo {author} {\bibfnamefont {D.}~\bibnamefont
  {Massart}},\ }\bibfield  {title} {\bibinfo {title} {{Automatic program for
  peak detection and deconvolution of multi-overlapped chromatographic signals:
  Part I: Peak detection}},\ }\href@noop {} {\bibfield  {journal} {\bibinfo
  {journal} {J. Chromatogr. A}\ }\textbf {\bibinfo {volume} {1096}},\ \bibinfo
  {pages} {133} (\bibinfo {year} {2005})}\BibitemShut {NoStop}%
\bibitem [{\citenamefont {O'Haver}\ \emph {et~al.}(1982)\citenamefont
  {O'Haver}, \citenamefont {Fell}, \citenamefont {Smith}, \citenamefont {Gans},
  \citenamefont {Sneddon}, \citenamefont {Bezur}, \citenamefont {Michel},
  \citenamefont {Ottaway}, \citenamefont {Miller}, \citenamefont {Ahmad},
  \citenamefont {Fell},\ and\ \citenamefont {andC.T. Cottrell}}]{OHaver1982}%
  \BibitemOpen
  \bibfield  {author} {\bibinfo {author} {\bibfnamefont {T.~C.}\ \bibnamefont
  {O'Haver}}, \bibinfo {author} {\bibfnamefont {A.~F.}\ \bibnamefont {Fell}},
  \bibinfo {author} {\bibfnamefont {G.}~\bibnamefont {Smith}}, \bibinfo
  {author} {\bibfnamefont {P.}~\bibnamefont {Gans}}, \bibinfo {author}
  {\bibfnamefont {J.}~\bibnamefont {Sneddon}}, \bibinfo {author} {\bibfnamefont
  {L.}~\bibnamefont {Bezur}}, \bibinfo {author} {\bibfnamefont {R.~G.}\
  \bibnamefont {Michel}}, \bibinfo {author} {\bibfnamefont {J.~M.}\
  \bibnamefont {Ottaway}}, \bibinfo {author} {\bibfnamefont {J.~N.}\
  \bibnamefont {Miller}}, \bibinfo {author} {\bibfnamefont {T.~A.}\
  \bibnamefont {Ahmad}}, \bibinfo {author} {\bibfnamefont {A.~F.}\ \bibnamefont
  {Fell}},\ and\ \bibinfo {author} {\bibfnamefont {B.~P.~C.}\ \bibnamefont
  {andC.T. Cottrell}},\ }\bibfield  {title} {\bibinfo {title} {{Derivative
  spectroscopy and its applications in analysis}},\ }\href@noop {} {\bibfield
  {journal} {\bibinfo  {journal} {Anal. Proc.}\ }\textbf {\bibinfo {volume}
  {19}},\ \bibinfo {pages} {22} (\bibinfo {year} {1982})}\BibitemShut {NoStop}%
\bibitem [{\citenamefont {Aspelmeyer}\ \emph {et~al.}(2014)\citenamefont
  {Aspelmeyer}, \citenamefont {Kippenberg},\ and\ \citenamefont
  {Marquardt}}]{Aspelmeyer2014}%
  \BibitemOpen
  \bibfield  {author} {\bibinfo {author} {\bibfnamefont {M.}~\bibnamefont
  {Aspelmeyer}}, \bibinfo {author} {\bibfnamefont {T.~J.}\ \bibnamefont
  {Kippenberg}},\ and\ \bibinfo {author} {\bibfnamefont {F.}~\bibnamefont
  {Marquardt}},\ }\bibfield  {title} {\bibinfo {title} {{Cavity
  Optomechanics}},\ }\href@noop {} {\bibfield  {journal} {\bibinfo  {journal}
  {Rev. Mod. Phys.}\ }\textbf {\bibinfo {volume} {86}},\ \bibinfo {pages}
  {1391} (\bibinfo {year} {2014})}\BibitemShut {NoStop}%
\bibitem [{\citenamefont {Stillinger}\ and\ \citenamefont
  {Herrick}(1975)}]{Stilinger1975}%
  \BibitemOpen
  \bibfield  {author} {\bibinfo {author} {\bibfnamefont {F.~H.}\ \bibnamefont
  {Stillinger}}\ and\ \bibinfo {author} {\bibfnamefont {D.~R.}\ \bibnamefont
  {Herrick}},\ }\bibfield  {title} {\bibinfo {title} {{Bound states in the
  continuum}},\ }\href@noop {} {\bibfield  {journal} {\bibinfo  {journal}
  {Phys. Rev. A}\ }\textbf {\bibinfo {volume} {11}},\ \bibinfo {pages} {446}
  (\bibinfo {year} {1975})}\BibitemShut {NoStop}%
\bibitem [{\citenamefont {Hsu}\ \emph {et~al.}(2016)\citenamefont {Hsu},
  \citenamefont {Zhen}, \citenamefont {Stone}, \citenamefont {Joannopoulos},\
  and\ \citenamefont {Solja\v{c}ic}}]{Hsu2016}%
  \BibitemOpen
  \bibfield  {author} {\bibinfo {author} {\bibfnamefont {C.~W.}\ \bibnamefont
  {Hsu}}, \bibinfo {author} {\bibfnamefont {B.}~\bibnamefont {Zhen}}, \bibinfo
  {author} {\bibfnamefont {A.~D.}\ \bibnamefont {Stone}}, \bibinfo {author}
  {\bibfnamefont {J.~D.}\ \bibnamefont {Joannopoulos}},\ and\ \bibinfo {author}
  {\bibfnamefont {M.}~\bibnamefont {Solja\v{c}ic}},\ }\bibfield  {title}
  {\bibinfo {title} {{Bound states in the continuum}},\ }\href@noop {}
  {\bibfield  {journal} {\bibinfo  {journal} {Nat. Rev. Mat.}\ }\textbf
  {\bibinfo {volume} {1}},\ \bibinfo {pages} {16048} (\bibinfo {year}
  {2016})}\BibitemShut {NoStop}%
\bibitem [{\citenamefont {Marinica}\ \emph {et~al.}(2008)\citenamefont
  {Marinica}, \citenamefont {Borisov},\ and\ \citenamefont
  {Shabanov}}]{Marinica2008}%
  \BibitemOpen
  \bibfield  {author} {\bibinfo {author} {\bibfnamefont {D.~C.}\ \bibnamefont
  {Marinica}}, \bibinfo {author} {\bibfnamefont {A.~G.}\ \bibnamefont
  {Borisov}},\ and\ \bibinfo {author} {\bibfnamefont {S.~V.}\ \bibnamefont
  {Shabanov}},\ }\bibfield  {title} {\bibinfo {title} {{Bound States in the
  Continuum in Photonics}},\ }\href@noop {} {\bibfield  {journal} {\bibinfo
  {journal} {Phys. Rev. Lett.}\ }\textbf {\bibinfo {volume} {100}},\ \bibinfo
  {pages} {183902} (\bibinfo {year} {2008})}\BibitemShut {NoStop}%
\bibitem [{\citenamefont {Plotnik}\ \emph {et~al.}(2011)\citenamefont
  {Plotnik}, \citenamefont {Peleg}, \citenamefont {Dreisow}, \citenamefont
  {Heinrich}, \citenamefont {Nolte}, \citenamefont {Szameit},\ and\
  \citenamefont {Segev}}]{Plotnik2011}%
  \BibitemOpen
  \bibfield  {author} {\bibinfo {author} {\bibfnamefont {Y.}~\bibnamefont
  {Plotnik}}, \bibinfo {author} {\bibfnamefont {O.}~\bibnamefont {Peleg}},
  \bibinfo {author} {\bibfnamefont {F.}~\bibnamefont {Dreisow}}, \bibinfo
  {author} {\bibfnamefont {M.}~\bibnamefont {Heinrich}}, \bibinfo {author}
  {\bibfnamefont {S.}~\bibnamefont {Nolte}}, \bibinfo {author} {\bibfnamefont
  {A.}~\bibnamefont {Szameit}},\ and\ \bibinfo {author} {\bibfnamefont
  {M.}~\bibnamefont {Segev}},\ }\bibfield  {title} {\bibinfo {title}
  {{Experimental Observation of Optical Bound States in the Continuum}},\
  }\href@noop {} {\bibfield  {journal} {\bibinfo  {journal} {Phys. Rev. Lett.}\
  }\textbf {\bibinfo {volume} {107}},\ \bibinfo {pages} {183901} (\bibinfo
  {year} {2011})}\BibitemShut {NoStop}%
\bibitem [{\citenamefont {Molina}\ \emph {et~al.}(2012)\citenamefont {Molina},
  \citenamefont {Miroshnichenko},\ and\ \citenamefont {Kivshar}}]{Molina2012}%
  \BibitemOpen
  \bibfield  {author} {\bibinfo {author} {\bibfnamefont {M.~I.}\ \bibnamefont
  {Molina}}, \bibinfo {author} {\bibfnamefont {A.~E.}\ \bibnamefont
  {Miroshnichenko}},\ and\ \bibinfo {author} {\bibfnamefont {Y.~S.}\
  \bibnamefont {Kivshar}},\ }\bibfield  {title} {\bibinfo {title} {{Surface
  Bound States in the Continuum}},\ }\href@noop {} {\bibfield  {journal}
  {\bibinfo  {journal} {Phys. Rev. Lett.}\ }\textbf {\bibinfo {volume} {108}},\
  \bibinfo {pages} {070401} (\bibinfo {year} {2012})}\BibitemShut {NoStop}%
\bibitem [{\citenamefont {Hsu}\ \emph {et~al.}(2013{\natexlab{a}})\citenamefont
  {Hsu}, \citenamefont {Zhen}, \citenamefont {Lee}, \citenamefont {Chua},
  \citenamefont {Johnson}, \citenamefont {Joannopoulos},\ and\ \citenamefont
  {Solja\v{c}ic}}]{Hsu2013}%
  \BibitemOpen
  \bibfield  {author} {\bibinfo {author} {\bibfnamefont {C.~W.}\ \bibnamefont
  {Hsu}}, \bibinfo {author} {\bibfnamefont {B.}~\bibnamefont {Zhen}}, \bibinfo
  {author} {\bibfnamefont {J.}~\bibnamefont {Lee}}, \bibinfo {author}
  {\bibfnamefont {S.-L.}\ \bibnamefont {Chua}}, \bibinfo {author}
  {\bibfnamefont {S.~G.}\ \bibnamefont {Johnson}}, \bibinfo {author}
  {\bibfnamefont {J.~D.}\ \bibnamefont {Joannopoulos}},\ and\ \bibinfo {author}
  {\bibfnamefont {M.}~\bibnamefont {Solja\v{c}ic}},\ }\bibfield  {title}
  {\bibinfo {title} {{Observation of trapped light within the radiation
  continuum}},\ }\href@noop {} {\bibfield  {journal} {\bibinfo  {journal}
  {Nature}\ }\textbf {\bibinfo {volume} {499}},\ \bibinfo {pages} {188}
  (\bibinfo {year} {2013}{\natexlab{a}})}\BibitemShut {NoStop}%
\bibitem [{\citenamefont {Hsu}\ \emph {et~al.}(2013{\natexlab{b}})\citenamefont
  {Hsu}, \citenamefont {Zhen}, \citenamefont {Chua}, \citenamefont {Johnson},
  \citenamefont {Joannopoulos},\ and\ \citenamefont
  {Solja\v{c}ic}}]{Hsu2013_2}%
  \BibitemOpen
  \bibfield  {author} {\bibinfo {author} {\bibfnamefont {C.~W.}\ \bibnamefont
  {Hsu}}, \bibinfo {author} {\bibfnamefont {B.}~\bibnamefont {Zhen}}, \bibinfo
  {author} {\bibfnamefont {S.-L.}\ \bibnamefont {Chua}}, \bibinfo {author}
  {\bibfnamefont {S.~G.}\ \bibnamefont {Johnson}}, \bibinfo {author}
  {\bibfnamefont {J.~D.}\ \bibnamefont {Joannopoulos}},\ and\ \bibinfo {author}
  {\bibfnamefont {M.}~\bibnamefont {Solja\v{c}ic}},\ }\bibfield  {title}
  {\bibinfo {title} {{Bloch surface eigenstates within the radiation
  continuum}},\ }\href@noop {} {\bibfield  {journal} {\bibinfo  {journal}
  {Light Sci. App.}\ }\textbf {\bibinfo {volume} {2}},\ \bibinfo {pages} {84}
  (\bibinfo {year} {2013}{\natexlab{b}})}\BibitemShut {NoStop}%
\bibitem [{\citenamefont {Zhen}\ \emph {et~al.}(2014)\citenamefont {Zhen},
  \citenamefont {Hsu}, \citenamefont {Lu}, \citenamefont {Stone},\ and\
  \citenamefont {Solja\v{c}ic}}]{Zhen2014}%
  \BibitemOpen
  \bibfield  {author} {\bibinfo {author} {\bibfnamefont {B.}~\bibnamefont
  {Zhen}}, \bibinfo {author} {\bibfnamefont {C.~W.}\ \bibnamefont {Hsu}},
  \bibinfo {author} {\bibfnamefont {L.}~\bibnamefont {Lu}}, \bibinfo {author}
  {\bibfnamefont {A.~D.}\ \bibnamefont {Stone}},\ and\ \bibinfo {author}
  {\bibfnamefont {M.}~\bibnamefont {Solja\v{c}ic}},\ }\bibfield  {title}
  {\bibinfo {title} {{Topological Nature of Optical Bound States in the
  Continuum}},\ }\href@noop {} {\bibfield  {journal} {\bibinfo  {journal}
  {Phys. Rev. Lett.}\ }\textbf {\bibinfo {volume} {113}},\ \bibinfo {pages}
  {257401} (\bibinfo {year} {2014})}\BibitemShut {NoStop}%
\bibitem [{\citenamefont {Doeleman}\ \emph {et~al.}(2018)\citenamefont
  {Doeleman}, \citenamefont {Francesco~Monticone},\ and\ \citenamefont
  {Koenderink}}]{Doeleman2018}%
  \BibitemOpen
  \bibfield  {author} {\bibinfo {author} {\bibfnamefont {H.~M.}\ \bibnamefont
  {Doeleman}}, \bibinfo {author} {\bibfnamefont {A.~A.}\ \bibnamefont
  {Francesco~Monticone}, \bibfnamefont {Wouter den~Hollander}},\ and\ \bibinfo
  {author} {\bibfnamefont {A.~F.}\ \bibnamefont {Koenderink}},\ }\bibfield
  {title} {\bibinfo {title} {{Experimental observation of a polarization vortex
  at an optical bound state in the continuum}},\ }\href@noop {} {\bibfield
  {journal} {\bibinfo  {journal} {Nat. Phot.}\ }\textbf {\bibinfo {volume}
  {12}},\ \bibinfo {pages} {397} (\bibinfo {year} {2018})}\BibitemShut
  {NoStop}%
\bibitem [{\citenamefont {Akahane}\ \emph {et~al.}(2003)\citenamefont
  {Akahane}, \citenamefont {Asano}, \citenamefont {Song},\ and\ \citenamefont
  {Noda}}]{Akahane2003}%
  \BibitemOpen
  \bibfield  {author} {\bibinfo {author} {\bibfnamefont {Y.}~\bibnamefont
  {Akahane}}, \bibinfo {author} {\bibfnamefont {T.}~\bibnamefont {Asano}},
  \bibinfo {author} {\bibfnamefont {B.-S.}\ \bibnamefont {Song}},\ and\
  \bibinfo {author} {\bibfnamefont {S.}~\bibnamefont {Noda}},\ }\bibfield
  {title} {\bibinfo {title} {{High-Q photonic nanocavity in a two-dimensional
  photonic crystal}},\ }\href@noop {} {\bibfield  {journal} {\bibinfo
  {journal} {Nature}\ }\textbf {\bibinfo {volume} {425}},\ \bibinfo {pages}
  {944} (\bibinfo {year} {2003})}\BibitemShut {NoStop}%
\bibitem [{\citenamefont {Srinivasan}\ \emph {et~al.}(2004)\citenamefont
  {Srinivasan}, \citenamefont {Barclay}, \citenamefont {Borselli},\ and\
  \citenamefont {Painter}}]{Srinivasan2004}%
  \BibitemOpen
  \bibfield  {author} {\bibinfo {author} {\bibfnamefont {K.}~\bibnamefont
  {Srinivasan}}, \bibinfo {author} {\bibfnamefont {P.~E.}\ \bibnamefont
  {Barclay}}, \bibinfo {author} {\bibfnamefont {M.}~\bibnamefont {Borselli}},\
  and\ \bibinfo {author} {\bibfnamefont {O.}~\bibnamefont {Painter}},\
  }\bibfield  {title} {\bibinfo {title} {{Optical-fiber-based measurement of an
  ultrasmall volume high-Q photonic crystal microcavity}},\ }\href@noop {}
  {\bibfield  {journal} {\bibinfo  {journal} {Phys. Rev. B}\ }\textbf {\bibinfo
  {volume} {70}},\ \bibinfo {pages} {081306(R)} (\bibinfo {year}
  {2004})}\BibitemShut {NoStop}%
\bibitem [{\citenamefont {Armani}\ \emph {et~al.}(2003)\citenamefont {Armani},
  \citenamefont {Kippenberg}, \citenamefont {Spillane},\ and\ \citenamefont
  {Vahala}}]{Armani2003}%
  \BibitemOpen
  \bibfield  {author} {\bibinfo {author} {\bibfnamefont {D.~K.}\ \bibnamefont
  {Armani}}, \bibinfo {author} {\bibfnamefont {T.~J.}\ \bibnamefont
  {Kippenberg}}, \bibinfo {author} {\bibfnamefont {S.~M.}\ \bibnamefont
  {Spillane}},\ and\ \bibinfo {author} {\bibfnamefont {K.~J.}\ \bibnamefont
  {Vahala}},\ }\bibfield  {title} {\bibinfo {title} {{Ultra-high-Q toroid
  microcavity on a chip}},\ }\href@noop {} {\bibfield  {journal} {\bibinfo
  {journal} {Nature}\ }\textbf {\bibinfo {volume} {421}},\ \bibinfo {pages}
  {925} (\bibinfo {year} {2003})}\BibitemShut {NoStop}%
\bibitem [{\citenamefont {Galli}\ \emph {et~al.}(2009)\citenamefont {Galli},
  \citenamefont {Portalupi}, \citenamefont {Belotti}, \citenamefont {Andreani},
  \citenamefont {O’Faolain}, ,\ and\ \citenamefont {Krauss}}]{Galli2009}%
  \BibitemOpen
  \bibfield  {author} {\bibinfo {author} {\bibfnamefont {M.}~\bibnamefont
  {Galli}}, \bibinfo {author} {\bibfnamefont {S.~L.}\ \bibnamefont
  {Portalupi}}, \bibinfo {author} {\bibfnamefont {M.}~\bibnamefont {Belotti}},
  \bibinfo {author} {\bibfnamefont {L.~C.}\ \bibnamefont {Andreani}}, \bibinfo
  {author} {\bibfnamefont {L.}~\bibnamefont {O’Faolain}}, ,\ and\ \bibinfo
  {author} {\bibfnamefont {T.~F.}\ \bibnamefont {Krauss}},\ }\bibfield  {title}
  {\bibinfo {title} {{Light scattering and Fano resonances in high-Q photonic
  crystal nanocavities}},\ }\href@noop {} {\bibfield  {journal} {\bibinfo
  {journal} {Appl. Phys. Lett.}\ }\textbf {\bibinfo {volume} {94}},\ \bibinfo
  {pages} {071101} (\bibinfo {year} {2009})}\BibitemShut {NoStop}%
\bibitem [{\citenamefont {Leijssen}\ and\ \citenamefont
  {Verhagen}(2015)}]{Leijssen2015}%
  \BibitemOpen
  \bibfield  {author} {\bibinfo {author} {\bibfnamefont {R.}~\bibnamefont
  {Leijssen}}\ and\ \bibinfo {author} {\bibfnamefont {E.}~\bibnamefont
  {Verhagen}},\ }\bibfield  {title} {\bibinfo {title} {{Strong optomechanical
  interactions in a sliced photonic crystal nanobeam}},\ }\href@noop {}
  {\bibfield  {journal} {\bibinfo  {journal} {Sci. Rep.}\ }\textbf {\bibinfo
  {volume} {5}},\ \bibinfo {pages} {15974} (\bibinfo {year}
  {2015})}\BibitemShut {NoStop}%
\bibitem [{Note1()}]{Note1}%
  \BibitemOpen
  \bibinfo {note} {The vanishing intensity of the X-pol signal prevents its
  direct demodulation at the LIA, forcing the use of the stronger U-pol
  signal.}\BibitemShut {Stop}%
\bibitem [{\citenamefont {Baldacci}\ \emph {et~al.}(2016)\citenamefont
  {Baldacci}, \citenamefont {Pitanti}, \citenamefont {Masini}, \citenamefont
  {Arcangeli}, \citenamefont {Colangelo}, \citenamefont {Navarro-Urrios},\ and\
  \citenamefont {Tredicucci}}]{Baldacci2016}%
  \BibitemOpen
  \bibfield  {author} {\bibinfo {author} {\bibfnamefont {L.}~\bibnamefont
  {Baldacci}}, \bibinfo {author} {\bibfnamefont {A.}~\bibnamefont {Pitanti}},
  \bibinfo {author} {\bibfnamefont {L.}~\bibnamefont {Masini}}, \bibinfo
  {author} {\bibfnamefont {A.}~\bibnamefont {Arcangeli}}, \bibinfo {author}
  {\bibfnamefont {F.}~\bibnamefont {Colangelo}}, \bibinfo {author}
  {\bibfnamefont {D.}~\bibnamefont {Navarro-Urrios}},\ and\ \bibinfo {author}
  {\bibfnamefont {A.}~\bibnamefont {Tredicucci}},\ }\bibfield  {title}
  {\bibinfo {title} {{Thermal noise and optomechanical features in the emission
  of a membrane-coupled compound cavity laser diode}},\ }\href@noop {}
  {\bibfield  {journal} {\bibinfo  {journal} {Sci. Rep.}\ }\textbf {\bibinfo
  {volume} {6}},\ \bibinfo {pages} {31489} (\bibinfo {year}
  {2016})}\BibitemShut {NoStop}%
\bibitem [{\citenamefont {Zanotto}\ \emph {et~al.}(2019)\citenamefont
  {Zanotto}, \citenamefont {Tredicucci}, \citenamefont {Navarro-Urrios},
  \citenamefont {Cecchini}, \citenamefont {Biasiol}, \citenamefont
  {Mencarelli}, \citenamefont {Pierantoni},\ and\ \citenamefont
  {Pitanti}}]{Zanotto2019}%
  \BibitemOpen
  \bibfield  {author} {\bibinfo {author} {\bibfnamefont {S.}~\bibnamefont
  {Zanotto}}, \bibinfo {author} {\bibfnamefont {A.}~\bibnamefont {Tredicucci}},
  \bibinfo {author} {\bibfnamefont {D.}~\bibnamefont {Navarro-Urrios}},
  \bibinfo {author} {\bibfnamefont {M.}~\bibnamefont {Cecchini}}, \bibinfo
  {author} {\bibfnamefont {G.}~\bibnamefont {Biasiol}}, \bibinfo {author}
  {\bibfnamefont {D.}~\bibnamefont {Mencarelli}}, \bibinfo {author}
  {\bibfnamefont {L.}~\bibnamefont {Pierantoni}},\ and\ \bibinfo {author}
  {\bibfnamefont {A.}~\bibnamefont {Pitanti}},\ }\bibfield  {title} {\bibinfo
  {title} {{Optomechanics of Chiral Dielectric Metasurfaces}},\ }\href@noop {}
  {\bibfield  {journal} {\bibinfo  {journal} {Adv. Opt. Mat.}\ }\textbf
  {\bibinfo {volume} {8}},\ \bibinfo {pages} {1901507} (\bibinfo {year}
  {2019})}\BibitemShut {NoStop}%
\bibitem [{Note2()}]{Note2}%
  \BibitemOpen
  \bibinfo {note} {Q-factors are limited by molecular film damping, although
  their value do not directly impact the performance of OMS as long as the
  resonator is not overdamped (like at atmospheric pressure) and its spectral
  features can be clearly recognized.}\BibitemShut {Stop}%
\bibitem [{\citenamefont {Gomis-Bresco}\ \emph {et~al.}(2014)\citenamefont
  {Gomis-Bresco}, \citenamefont {Navarro-Urrios}, \citenamefont {Oudich},
  \citenamefont {El-Jallal}, \citenamefont {Griol}, \citenamefont {Puerto},
  \citenamefont {Chavez}, \citenamefont {Pennec}, \citenamefont
  {Djafari-Rouhani}, \citenamefont {Alzina}, \citenamefont {Martínez},\ and\
  \citenamefont {Sotomayor-Torres}}]{GomisBresco2014}%
  \BibitemOpen
  \bibfield  {author} {\bibinfo {author} {\bibfnamefont {J.}~\bibnamefont
  {Gomis-Bresco}}, \bibinfo {author} {\bibfnamefont {D.}~\bibnamefont
  {Navarro-Urrios}}, \bibinfo {author} {\bibfnamefont {M.}~\bibnamefont
  {Oudich}}, \bibinfo {author} {\bibfnamefont {S.}~\bibnamefont {El-Jallal}},
  \bibinfo {author} {\bibfnamefont {A.}~\bibnamefont {Griol}}, \bibinfo
  {author} {\bibfnamefont {D.}~\bibnamefont {Puerto}}, \bibinfo {author}
  {\bibfnamefont {E.}~\bibnamefont {Chavez}}, \bibinfo {author} {\bibfnamefont
  {Y.}~\bibnamefont {Pennec}}, \bibinfo {author} {\bibfnamefont
  {B.}~\bibnamefont {Djafari-Rouhani}}, \bibinfo {author} {\bibfnamefont
  {F.}~\bibnamefont {Alzina}}, \bibinfo {author} {\bibfnamefont
  {A.}~\bibnamefont {Martínez}},\ and\ \bibinfo {author} {\bibfnamefont
  {C.}~\bibnamefont {Sotomayor-Torres}},\ }\bibfield  {title} {\bibinfo {title}
  {{A one-dimensional optomechanical crystal with a complete phononic band
  gap}},\ }\href@noop {} {\bibfield  {journal} {\bibinfo  {journal} {Nat.
  Comm.}\ }\textbf {\bibinfo {volume} {5}},\ \bibinfo {pages} {4452} (\bibinfo
  {year} {2014})}\BibitemShut {NoStop}%
\bibitem [{Note3()}]{Note3}%
  \BibitemOpen
  \bibinfo {note} {To improve the visibilities at large angles, a complementary
  dataset to the one of Fig. \ref {fig:4}, with individually normalized
  spectra, has been reported in the SI - sec. IV.B.}\BibitemShut {Stop}%
\bibitem [{PPM()}]{PPML}%
  \BibitemOpen
  \href@noop {} {\bibinfo {title} {{Simone Zanotto} ppml - periodically
  patterned multi layer}},\ \bibinfo {howpublished}
  {\url{https://www.mathworks.com/matlabcentral/fileexchange/55401-ppml-periodically-patterned-multi-layer}},\
  \bibinfo {note} {2021}\BibitemShut {NoStop}%
\bibitem [{Note4()}]{Note4}%
  \BibitemOpen
  \bibinfo {note} {For improved clarity of the waterfall plot, each curve has
  bee offset by a factor $c\cdot \protect \qopname \relax o{tan}^{-1}(10\alpha
  )$ , where $c$ is a constant and $\alpha $ the angle of
  incidence}\BibitemShut {NoStop}%
\bibitem [{\citenamefont {Zanotto}(2018)}]{Zanotto2018}%
  \BibitemOpen
  \bibfield  {author} {\bibinfo {author} {\bibfnamefont {S.}~\bibnamefont
  {Zanotto}},\ }\bibfield  {title} {\bibinfo {title} {Weak coupling, strong
  coupling, critical coupling and fano resonances: A unifying vision},\ }in\
  \href@noop {} {\emph {\bibinfo {booktitle} {Fano Resonances in Optics and
  Microwaves}}},\ \bibinfo {editor} {edited by\ \bibinfo {editor}
  {\bibfnamefont {E.}~\bibnamefont {Kamenetskii}}, \bibinfo {editor}
  {\bibfnamefont {A.}~\bibnamefont {Sadreev}},\ and\ \bibinfo {editor}
  {\bibfnamefont {A.}~\bibnamefont {Miroshnichenko}}}\ (\bibinfo  {publisher}
  {Springer, Berlin},\ \bibinfo {year} {2018})\BibitemShut {NoStop}%
\bibitem [{\citenamefont {Fan}\ and\ \citenamefont
  {Joannopoulos}(2002)}]{Fan2002}%
  \BibitemOpen
  \bibfield  {author} {\bibinfo {author} {\bibfnamefont {S.}~\bibnamefont
  {Fan}}\ and\ \bibinfo {author} {\bibfnamefont {J.}~\bibnamefont
  {Joannopoulos}},\ }\bibfield  {title} {\bibinfo {title} {{Analysis of guided
  resonances in photonic crystal slabs}},\ }\href@noop {} {\bibfield  {journal}
  {\bibinfo  {journal} {Phys. Rev. B}\ }\textbf {\bibinfo {volume} {65}},\
  \bibinfo {pages} {235112} (\bibinfo {year} {2002})}\BibitemShut {NoStop}%
\bibitem [{\citenamefont {Fan}\ \emph {et~al.}(2003)\citenamefont {Fan},
  \citenamefont {Suh},\ and\ \citenamefont {Joannopoulos}}]{Fan2003}%
  \BibitemOpen
  \bibfield  {author} {\bibinfo {author} {\bibfnamefont {S.}~\bibnamefont
  {Fan}}, \bibinfo {author} {\bibfnamefont {W.}~\bibnamefont {Suh}},\ and\
  \bibinfo {author} {\bibfnamefont {J.}~\bibnamefont {Joannopoulos}},\
  }\bibfield  {title} {\bibinfo {title} {{Temporal coupled-mode theory for the
  Fano resonance in optical resonators}},\ }\href@noop {} {\bibfield  {journal}
  {\bibinfo  {journal} {J. Opt. Soc. Am. A}\ }\textbf {\bibinfo {volume}
  {20}},\ \bibinfo {pages} {569} (\bibinfo {year} {2003})}\BibitemShut
  {NoStop}%
\bibitem [{\citenamefont {R.Ohta}\ \emph {et~al.}(2018)\citenamefont {R.Ohta},
  \citenamefont {Okamoto}, \citenamefont {Tawara}, \citenamefont {Gotoh},\ and\
  \citenamefont {Yamaguchi}}]{Ohta2018}%
  \BibitemOpen
  \bibfield  {author} {\bibinfo {author} {\bibnamefont {R.Ohta}}, \bibinfo
  {author} {\bibfnamefont {H.}~\bibnamefont {Okamoto}}, \bibinfo {author}
  {\bibfnamefont {T.}~\bibnamefont {Tawara}}, \bibinfo {author} {\bibfnamefont
  {H.}~\bibnamefont {Gotoh}},\ and\ \bibinfo {author} {\bibfnamefont
  {H.}~\bibnamefont {Yamaguchi}},\ }\bibfield  {title} {\bibinfo {title}
  {{Dynamic Control of the Coupling between Dark and Bright Excitons with
  Vibrational Strain}},\ }\href@noop {} {\bibfield  {journal} {\bibinfo
  {journal} {Phys. Rev. Lett.}\ }\textbf {\bibinfo {volume} {120}},\ \bibinfo
  {pages} {267401} (\bibinfo {year} {2018})}\BibitemShut {NoStop}%
\end{thebibliography}%

\end{document}